\documentclass[aps,prl,twocolumn,groupedaddress,showpacs]{revtex4}

\usepackage{graphicx}

\newcommand{\rydn}{\textsf{n}}

\begin{document}


\title{Mechanical effect of van der Waals interactions \\
 observed in real time in an ultracold Rydberg gas}


\author{T. Amthor}
\email[]{thomas.amthor@physik.uni-freiburg.de}
\author{M. Reetz-Lamour}
\author{S. Westermann}
\author{J. Denskat}
\author{M. Weidem\"uller}
\email[]{m.weidemueller@physik.uni-freiburg.de}
\affiliation{Physikalisches Institut, Universit\"at Freiburg,
  Hermann-Herder-Str. 3, 79104 Freiburg, Germany}


\date{\today}

\begin{abstract}
We present time-resolved spectroscopic measurements of
Rydberg-Rydberg interactions in an ultracold gas, revealing the pair
dynamics induced by long-range van der Waals interactions between
the atoms. By detuning the excitation laser, a specific pair distribution 
is prepared. Penning ionization on a
microsecond timescale serves as a probe for the pair dynamics under
the influence of the attractive long-range forces. Comparison with a
Monte Carlo model not only explains all spectroscopic features but
also gives quantitative information about the interaction
potentials. The results imply that the interaction-induced
ionization rate can be influenced by the excitation laser.
Surprisingly, interaction-induced ionization is also observed for Rydberg
states with purely repulsive interactions.
\end{abstract}


\maketitle

Long-range dipolar interactions ubiquitously appear in nature as the
cause for binding forces ranging from atomic and molecular gases
\cite{santos00} all the way to large biological systems
\cite{autumn00}. Rydberg atoms have attracted much interest in this
context, as they represent an ideal system to study quantum dynamics
under the influence of dipolar interactions. As a
prominent example from cavity quantum electrodynamics, attractive
forces between Rydberg atoms and conducting surfaces have been
observed as level shifts in atomic beam experiments
\cite{sandoghdar92}.
Rydberg-Rydberg interactions leading to resonant energy transfer
have been studied in thermal beam experiments \cite{safinya81} and
later in ultracold Rydberg gases \cite{anderson98,mourachko98,westermann06}.
The long-range dipolar interactions can be used to block multiple
Rydberg excitation \cite{tong04,singer04} and to create many-particle
entangled states, which may be employed for quantum information
processing \cite{jaksch00,lukin01}. So far, investigations of
Rydberg-Rydberg interactions have mainly focused on the electronic
degrees of freedom neglecting the center-of-mass motion (``frozen
Rydberg gas'' \cite{mourachko98}). In this Letter, we present real-time
measurements of the motion of interacting pairs of Rydberg atoms
revealing the character and strength of the long-range interparticle
interactions.

In most molecular and atomic systems the relevant timescales of
interparticle dynamics are in the sub-ns regime calling for very
fast probes. Ultracold atoms bridge to the ns regime as the thermal
energy ($T$\textless1\,mK) is negligible which allows one to study
systems with much weaker interactions or large interatomic
distances. Due to the negligible kinetic energy, the dynamics of the
gas is fully determined by the interatomic interactions.
Interactions in a cold atomic sample have been studied time-resolved
in the case of ground state atoms using a pump-probe scheme
\cite{gensemer98} allowing for coherent control of the collision
process \cite{wright05}. Evidence for interaction-induced motion in
cold Rydberg gases was recently found spectroscopically as the cause
for Penning ionization \cite{li05}. By combining time-resolved and
spectroscopic measurements, we quantitatively examine the
van der Waals (vdW) interactions between two Rydberg atoms. Our
measurements can be directly compared to calculations of
induced-dipole interaction potentials for two Rydberg atoms, which
have so far only been performed neglecting spin-orbit coupling
\cite{singer05b}. These interaction potentials may even
bear bound states resulting in the formation of exotic
ultralong-range molecules \cite{boisseau02}.

\begin{figure}
 \includegraphics{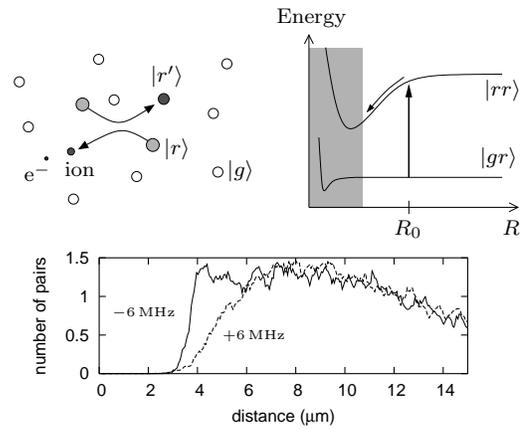}
 \caption{\label{fig:levelscheme}Excitation and interaction-induced ionization
  of a pair of atoms on an attractive vdW potential. The atoms are
  initially excited to state $|r\rangle$ at a distance $R_0$ by a
  slightly red-detuned laser. After excitation
  they are accelerated towards each other and collide, leading to ionization
  of one of the atoms. The region where
  this Penning ionization can occur is indicated in gray.
  The number of pairs excited at specific distances is shown in the lower
  graph for different detunings.}
\end{figure}

The idea of the experiment is schematically illustrated in Fig.
\ref{fig:levelscheme}. The vdW interaction between two Rydberg atoms
leads to an energy shift of the pair excitation which depends on the
pair distance. A red-detuned laser will therefore preferably excite
atoms on an attractive potential in a certain range of distances
around an already excited atom. In this way one can use the laser
detuning to create a sample with a specific distribution of pair
distances. Two examples for $+6$\,MHz (blue) and $-6$\,MHz (red)
detuning as derived from our model are also plotted in Fig.
\ref{fig:levelscheme}. Once a pair of Rydberg atoms is excited on an
attractive potential, the atoms will be accelerated towards each
other and collide after a certain time depending on their initial
distance $R_0$. These collisions can lead to Penning ionization.
By measuring the number of ions produced
after a variable time $\Delta t$ one can follow the dynamics of the system in
real time. The ionization is thus used as a monitor signal for pair dynamics
giving quantitative information about the interaction strength.

In our setup we trap $^{87}$Rb atoms at temperatures below 100\,$\mu$K and a
peak density on the order of $10^{10}$cm$^{-3}$ in a magneto-optical trap
(MOT). The atoms are then excited to Rydberg states
using a two-photon excitation scheme. The two atomic transitions
5S$_{1/2}\rightarrow$ 5P$_{3/2}$ and 5P$_{3/2}\rightarrow \rydn\ell$ are
realized with two cw laser systems at 780\,nm and 480\,nm, respectively.
The frequency of the 480-nm excitation laser is actively
stabilized using an ultra-stable reference cavity. 
The 480-nm laser is focused to a waist of $\sim 37\,\mu$m at the
center of the MOT with a power of 10\,mW.
Two metal grids are used
to apply electric fields for stray field compensation and for field ionization
of the Rydberg states. Ions are detected on a microchannel plate detector.
Ref. \cite{singer05a} describes the setup in more detail.

The experimental cycle, repeated every 70\,ms, is as follows: The
excitation laser is switched on for 100\,ns at a given detuning.
This time has been
chosen sufficiently short so that the movement of the atoms
is negligible and no ionization takes place during the excitation.
The gas can then evolve freely for a variable time $\Delta t$. After that,
an electric field ramp is applied to field ionize the Rydberg atoms and
accelerate the ions towards the detector. Ions produced by collisions will be
drawn to the detector at the very beginning of the ramp (i.e. at low
accelerating fields), while Rydberg
states ionize at a finite electric field, generating a delayed detector signal.
Using two boxcar integrators, the two signals are recorded simultaneously.

\begin{figure}
 \includegraphics[scale=1.]{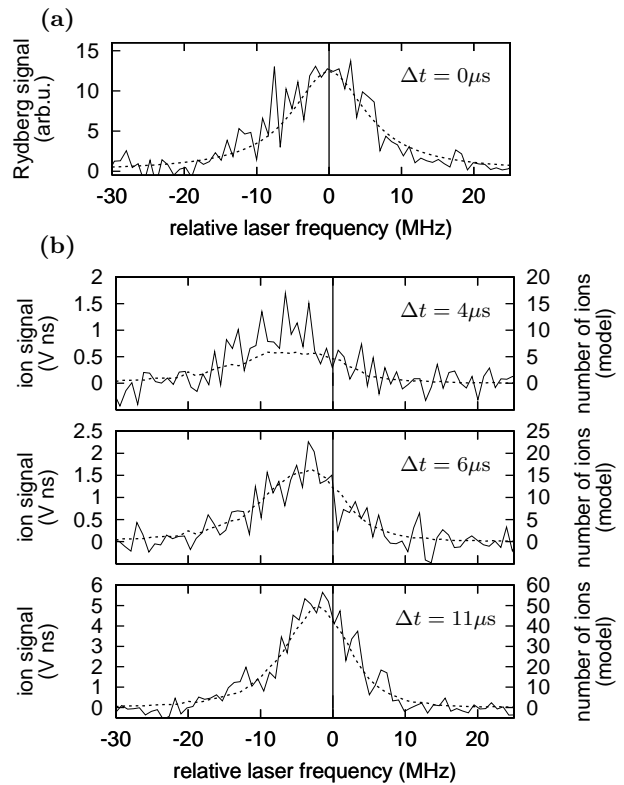}
 \caption{\label{fig:montecarlo}
  (a)  60D$_{5/2}$ Rydberg excitation line (solid), with fitted
  Lorentzian (dotted),
  (b) development of the
  ion signal for different interaction times (solid) compared to
  Monte-Carlo simulation (dotted).
  The laser frequency is given relative to the atomic resonance.}
\end{figure}

Fig. \ref{fig:montecarlo} shows the Rydberg and ion signals at the 60D$_{5/2}$
resonance as a function of the excitation laser frequency. The frequency axis
is centered at the atomic resonance.
In Fig. \ref{fig:montecarlo}a the Rydberg signal measured directly after
excitation is displayed. The width of the excitation line is given by
the width of the intermediate 5P level (6\,MHz FWHM), the laser linewidths
(2-5\,MHz FWHM), the finite excitation time (resulting in a
4.2\,MHz FWHM of the Fourier transform), and a $\sim$2\,MHz
broadening due to residual electric fields.
The line shape should thus be a Voigt profile, but it is still described
reasonably well by a Lorentz fit (also shown in the graph) yielding a FWHM of
12.5\,MHz as expected from the above numbers.
In the graphs in Fig. \ref{fig:montecarlo}b the development of the ion signal
after different interaction times $\Delta t$ is shown. All data are compared to
the results of the simulation (dotted lines) described below.
Within several $\mu$s the number of ions increases steadily. The
ion signal peaks at the red-detuned side of the atomic resonance
and exhibits a pronounced red wing.
As the interaction time increases, most ions appear nearer to the atomic
resonance, which can be observed as a shift of the ionization line towards
zero detuning.
These spectral features can easily be understood in the picture of
colliding pairs: After a short interaction time only very close pairs will
have had the time to collide, and these pairs are preferably excited at large
detuning.  For very long times, the line shape resembles the initial Rydberg
excitation line, as almost all atoms are ionized.
From these results it is obvious that by slight detuning of the excitation
laser the initial ionization rate can be controlled.
This can be of importance for applications like
quantum computation with Rydberg atoms, where the presence of ions
acts as a decoherence process, and for the investigation of ultracold plasma
formation.

While a simple analytical description of pair excitations as
in Ref. \cite{galpritch89} can qualitatively explain the behavior,
it turns out to be insufficient to describe the complex excitation
process and thus does not provide quantitative information about the
interaction potentials. Instead of considering only the excitation
of individual pairs far away from each other our model accounts for
the interaction between all atoms and allows for clustering of
Rydberg excitation around already excited atoms in order to describe
the system more realistically and to reproduce the measured line
shapes. We have therefore performed a Monte Carlo simulation.
100\,000 atoms are randomly placed in a box with a Gaussian
distribution along the excitation laser axis to represent the MOT
with a peak density of $1\times 10^{10}$\,cm$^{-3}$. Excitation of
the gas is described by an iterative procedure: In each iteration
step, each of the atoms can be excited with a certain probability.
For atom $i$ it is given by
$ P_i \propto \mathcal L(\delta-V(R_i))\,\exp(-2(x_i^2+y_i^2)/w^2)$.
$\mathcal L$ is the line profile representing the 12.5\,MHz linewidth of the
Rydberg excitation as described above.
$\mathcal L$ is evaluated at a detuning consisting of the laser detuning
$\delta$ and the interaction energy $V(R_i)$
with the nearest Rydberg neighbor to atom $i$.
The Gaussian intensity distribution of the excitation laser with waist
$w=37\,\mu$m is accounted for by the last factor.
After each Rydberg excitation, the table of nearest Rydberg neighbors is
recalculated for all atoms.
This excitation process is iterated to yield the actual excitation fraction of
a few percent at the center of the MOT.

The excited atoms are then combined to disjoint
pairs by successively choosing the two atoms closest to each other.
From the resulting list of distances, shown as a histogram in
the lower graph in Fig. \ref{fig:levelscheme}, the number of
pairs colliding until a certain time $\Delta t$ can be determined.
We assume an attractive potential of the form $V(R)=-C_6/R^6$. The time it
takes a pair of atoms with initial distance $R_0$ and an initial velocity of
zero to collide can then be approximated by the time to reach a distance of
zero, 
$ \tau_{\mathrm coll}\approx 0.2156\,R_0^4\,(2\,C_6/m_\mathrm{Rb})^{-1/2} $
\cite{galpritch89}. To determine the number of ions produced until a time
$\Delta t$, we count the pairs that collide within this time. The whole
simulation is performed for different laser detunings to
obtain the line shape of the ion signal.

The resulting traces shown in Fig. \ref{fig:montecarlo} are averaged over ten
random realizations as described above.
The best agreement of the Monte Carlo simulation with the experimental results
is found for $C_6=2\times 10^{20}$\,au, while values below $10^{20}$
and above $10^{21}$ are not compatible with the measured spectra.
This is in agreement with calculations following
Ref. \cite{singer05b}. Note that in the actual experiment an effective
potential is observed, which averages over all possible molecular
symmetries.
The measured ionization signals (Fig. \ref{fig:montecarlo}b) are well
reproduced by the simulation.

\begin{figure}
 \includegraphics{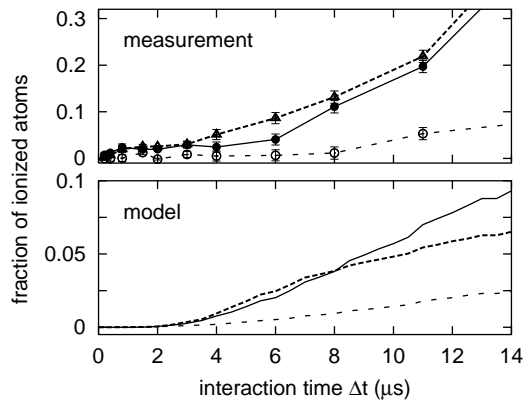}
\caption{\label{fig:timesingle}Signal growth at different detunings: $0$\,MHz
 (solid), $+6$\,MHz (dotted), $-6$\,MHz (dashed).
 Upper graph: measured data (data points averaged over 3\,MHz), lower graph:
  model. The scale of the measurement data contains a systematic error of a
 factor of two. }
\end{figure}

While this spectroscopic view shows how the line shape is reproduced by the
simulation, one can gain some more insight in the system by comparing the time
development of the measured signal and the model at specific detunings, as
displayed in Fig. \ref{fig:timesingle}.
The model predicts a threshold for the ion production at around
3-4\,$\mu$s, which is also observed in the experiment. Furthermore, when
comparing the ionization rates at different detunings relative to each other,
we find very similar behavior of model and experiment: At $-6$\,MHz the
rate is comparable to the one at zero detuning, while at $+6$\,MHz the rate is
significantly lower. The error bars at the measured data points display the
standard deviation of fluctuations in the signal, while the scale is subject
to a systematic error of a factor of two.

So far we have considered the interaction potential to be of pure vdW
type. However, as pointed out in Ref. \cite{li05}, nearby dipole-coupled pair
states introduce a $-1/R^3$ behavior of the interaction potential for small
distances.
To estimate the influence of this effect we have used a simple interaction
Hamiltonian $\mathcal H$ considering only the strongest couplings.
From our Monte Carlo model we infer that the distance between pairs of atoms is
always above 3\,$\mu$m (see Fig. \ref{fig:levelscheme}).
These distances are above the vdW radius (the distance at which the
$-1/R^3$ behavior becomes apparent), {\it i.e.} the eigenenergies of
$\mathcal H$ do not differ significantly from a pure $-C_6/R^6$ potential with
$C_6=5.4\times 10^{20}$\,au.
Even when comparing the collision times of pairs
derived from these potentials we find only slight differences for typical
initial distances. We thus expect a description in terms of a vdW
potential to be appropriate.
In contrast, when re-running the Monte Carlo calculations considering only the
$-1/R^3$ component of the potential, the ion signal line shape could not be
reproduced at all.

This description will only be applicable as long as cold Rydberg-Rydberg
collisions are the main cause of ionization. As soon as the ion density
becomes sufficiently large to trap electrons, electron-Rydberg collisions
will determine the dynamics of the system, causing ionization as well as
$\ell$ and \rydn~mixing \cite{walz04}.
Redistribution of atoms to other Rydberg states by
Penning ionization or by black-body radiation (estimated rates below
5600\,s$^{-1}$ \cite{gallagher94}) as well as direct black-body ionization
will also affect the dynamics of the system, but should not have significant
influence during the first few $\mu$s.
For long interaction times, however, we expect our model to become increasingly
inaccurate and the measured signal to exceed the simulated one. 
This is indeed what we observe for $\Delta t>11\,\mu$s (see
Fig. \ref{fig:timesingle}).

\begin{figure}
 \includegraphics{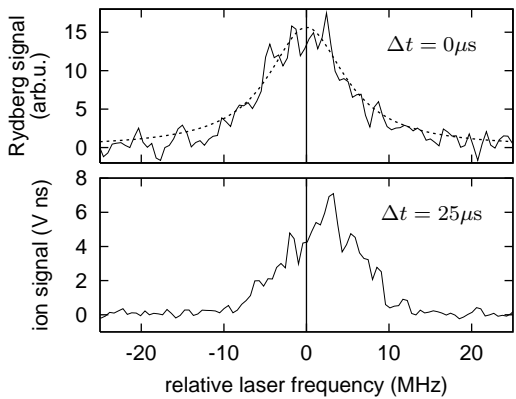}
 \caption{\label{fig:rydandionS}60S Rydberg excitation line (upper graph) and
  pre-ionized atoms after 25$\mu$s interaction time (lower graph).
  The ions are produced on the blue-detuned side of the resonance.
  Excitation time is 300\,ns, and excitation laser power 22\,mW.
  The graphs have different signal height scales. The fit in the upper graph
  (dotted) yields a line width of 11.3\,MHz, similar to the 60D$_{5/2}$ line.}
\end{figure}

According to the above analysis one would not expect to see ionization due to
inelastic processes when the interaction is repulsive.
We have also performed similar measurements for the 60S state,
which exhibits repulsive vdW interaction \cite{singer05b}.
Surprisingly, we still observe ionization, yet only on much longer timescales.
Most ions are produced on the blue-detuned side of the resonance,
which implies that the ionization rate is in conjuction with the repulsive
potential.
A possible explanation is the redistribution of states due to black-body
radiation leading to attractive dipole-dipole potentials. Atoms initially
excited at small distance ({\it i.e.} at blue detuning) could then be
the first to collide \cite{gallagherpc}.
This process is more difficult to model for a cloud of atoms and a simple
approach of including black-body redistribution and subsequent $1/R^3$
attraction in our Monte Carlo model could not explain the observed line
shapes. There may also exist attractive pair potentials of
other states crossing the repulsive vdW potentials.
Atoms initially repelled from each other may reach a crossing and
return on the attractive path.
We will further investigate this process in future experiments.

In conclusion, the time-resolved spectroscopic measurements reveal
the motion of Rydberg atoms under the influence of van der Waals
interactions as the cause of Penning ionization. The temporal
evolution of the ionization signal allows one to distinguish
van der Waals from resonant dipole interactions. The measured
strength of the interaction is in good quantitative agreement with
theoretical predictions. 
The accuracy of determining the interaction strength could be increased
by using two excitation laser pulses at
different frequencies: The first one to excite a dilute Rydberg gas
on the atomic resonance,
the second one detuned from resonance to
create well-defined pair distributions over a larger range of accessible
detunings. 
F\"orster resonances may be
used to switch the character of the interaction from van der Waals
to resonant dipole interaction. In this way our experimental method
will allow for a more precise determination of long-range potentials
and challenge refined models of many-body dynamics under the
influence of dipolar forces.

\begin{acknowledgments}
The project is supported in part by the Landes\-stiftung
Baden-W\"urttemberg in the framework of the ``Quantum Information
Processing'' program, and a grant from the Ministry of Science,
Research and Arts of Baden-W\"urttemberg (Az: 24-7532.23-11-11/1).
We thank T. F. Gallagher and P. Gould for inspiring discussions.
\end{acknowledgments}


\end{document}